\documentclass[a4paper,fleqn]{cas-dc}
\usepackage[utf8]{inputenc}
\usepackage{geometry}
\usepackage{hyperref}
\hypersetup{
colorlinks= true,
urlcolor = blue,
linkcolor = red,
citecolor = blue,
}
\usepackage{todonotes}
\usepackage[numbers]{natbib}
\usepackage{bigstrut}

\begin{document}
\let\WriteBookmarks\relax
\def\floatpagepagefraction{1}
\def\textpagefraction{.001}
\shorttitle{Artificial Intelligence for In Silico Clinical Trials}
\shortauthors{Wang et~al.}

\title[mode = title]{Artificial Intelligence for In Silico Clinical Trials: A Review}

\author[1]{Zifeng Wang}
\fnmark[1]

\ead{zifengw2@illinois.edu}

\author[1]{Chufan Gao}
\fnmark[1]
\ead{chufan2@illinois.edu}

\author[3]{Lucas M. Glass}

\author[1,2]{Jimeng Sun}[orcid=0000-0003-1512-6426]
\cormark[2]
\ead{jimeng@illinois.edu}

\address[1]{Department of Computer Science, University of Illinois Urbana-Champaign}
\address[2]{Carle Illinois College of Medicine, University of Illinois Urbana-Champaign}
\address[3]{Analytics Center of Excellence, IQVIA}

\begin{abstract}
A clinical trial is an essential step in drug development, which is often costly and time-consuming. In silico trials are clinical trials conducted digitally through simulation and modeling as an alternative to traditional clinical trials. 
AI-enabled in silico trials can increase the case group size by creating virtual cohorts as controls. In addition, it also enables automation and optimization of trial design  and predicts the trial success rate. This article systematically reviews papers under three main topics: clinical simulation, individualized predictive modeling, and computer-aided trial design. We focus on how machine learning (ML) may be applied in these applications. In particular, we present the machine learning problem formulation and available data sources for each task. We end with discussing the challenges and opportunities of AI for in silico trials in real-world applications.
\end{abstract}


\maketitle

\footnotetext[1]{This author contributed equally to the first author.}

\section{Introduction}

A clinical trial is a prospective study aiming to compare the effects and value of new interventions in human subjects. 
The median cost of a clinical trial was US\$3.4 million for a phase I trial, \$8.6 million for a phase II trial, and \$21.4 million for a phase III trial~\cite{martin2017much}. 
The high cost comes from different activities such as protocol design, site identification, participant recruitment, treatment, and follow-up tracking. It is of broad interest to reduce the time and  cost of drug development and minimize the risks of adverse side effects.

In silico trials, also known as virtual trials, are clinical trials conducted digitally, in contrast to in vivo trials conducted in living organisms such as the human body. One of the capabilities of in silico trials is to develop individualized virtual cohorts for testing the safety and efficacy of new treatments through computer simulation, and modeling \cite{pappalardo2019silico}. In silico trials may also assist the design of clinical trials, thus facilitating more efficient trial execution. While other industries, such as manufacturing industries, have long used computer simulations to accelerate research and development for cars and buildings, the pharmaceutical industry has only recently started exploring in silico methods \cite{walsh2020generating}. This is probably due to the rapid growth of efficient, accurate, and reliable computational models and artificial intelligence (AI). 

Interest in AI for in silico trials has grown for two primary reasons. First, the rising prevalence of electronic health records (EHRs) and electronic data capturing systems (EDC)  have produced an unprecedented amount of patient data intractable for manual processing and analysis. In contrast, AI and machine learning (ML) models can take full advantage of patient data, automatically processing and mining patterns from such data. Second, AI has the potential to provide novel insights into many previously infeasible tasks, e.g., simulating digital twins for testing treatment efficacy. In light of this, this article summarizes the recent AI applications to in silico clinical trials and describes associated challenges and opportunities. 

\begin{figure*}[t]
\centering
\includegraphics[width=0.8\textwidth]{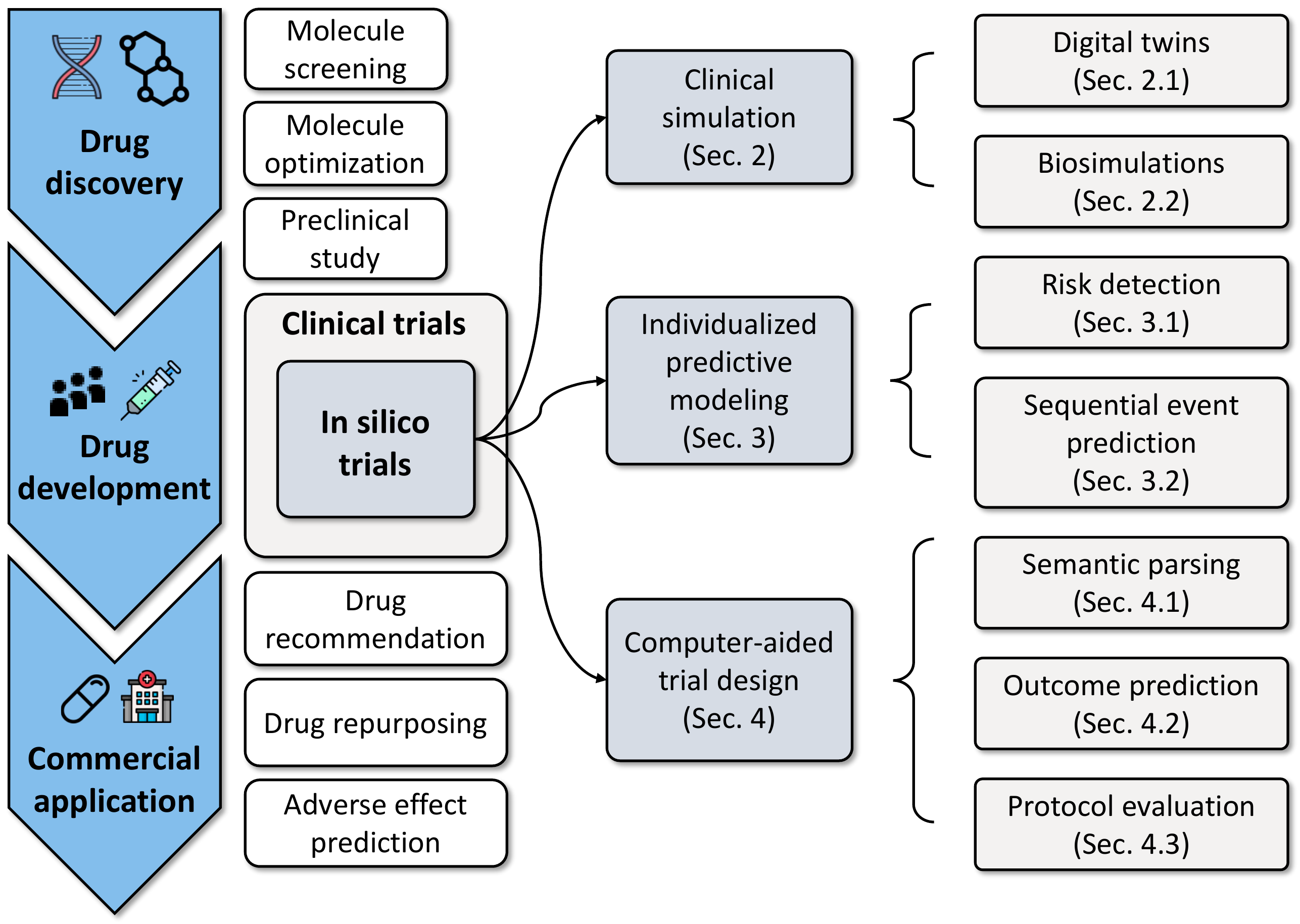}
\caption{Organization and coverage of this survey. This survey focuses on AI for clinical trials, emphasizing in silico clinical trials. We review a series of hot topics and then provide ML formulation and available data sources correspondingly. At last, we identify challenges and opportunities in this domain. \label{fig:overview}}
\end{figure*}

Fig. \ref{fig:overview} positions in silico trials  in the context of the entire pharmaceutical pipeline. In general, designing a drug undergoes two phases: drug discovery and drug development, before the new drug is commercialized into practice. In the discovery phase, compounds are first tested to modulate the activity of the biological target of interest through molecule screening. Then, the selected compounds will be optimized into lead compounds through molecule optimization and will be tested in animal models through preclinical studies. After that, the drug development phase begins with progressing the candidate drugs to approval through multi-phase human trials. Four phases of clinical trials can be performed in this stage, where in silico trials are instrumental. The data regarding the safety and efficacy of candidate drugs for human beings will be collected to support the approval for commercialization. Upon approval, post-market surveillance on  adverse effects across a large population can be conducted to monitor the safety of the approved drugs. Pharmaceutical companies also market new drugs to clinics through drug recommendation and consider drug repurposing to find new indications for existing drugs.

Several reviews of molecule design and genomics have been published \cite{huang2021machine,du2022molgensurvey}, which lie in the drug discovery space. There are also reviews of AI in clinical decision support and healthcare applications \cite{xiao2018opportunities,shahid2019applications}. However, we focus on the emerging area of in silico trials in the drug development stage. The most relevant works are \citet{pappalardo2019silico}, which introduced the advances of in silico trials, and \citet{weissler2021role}, which reviewed machine learning for general clinical research. However, we study AI tasks in the context of in silico trials, which are not covered in previous works.

In particular, we categorize AI applications of in silico trials into three main topics as shown by Fig. ~\ref{fig:overview}: 

\begin{enumerate}
    \item {\it Clinical simulation} is about creating virtual cohorts or organisms to study the efficacy and safety of treatments.
    \item {\it Individualized predictive modeling} is about building predictive models that adapt to personal characteristics to monitor and improve patient outcomes.
    \item {\it Computer-aided trial design} leverages ML and AI to organize, evaluate, and improve clinical trial protocols.
\end{enumerate}

We list the main technical papers mentioned in this article in Table \ref{tab:paper_list}, which will be discussed in later sections. In addition, we also present the machine learning formulations and available data sources for each task.  Since this review focuses on in silico trials, we will not cover how machine learning is used for other tasks in clinical trials, e.g., participant recruitment \cite{gao2020compose}, decentralized trials \cite{rieke2020future}, and trial site selection \cite{srinivasa2022clinical}.

\begin{table*}[t]
  \centering
  \caption{A list of papers related to AI and in silico clinical trials.}
     \resizebox{\textwidth}{!}{%
    \begin{tabular}{lllp{26.715em}}
    \hline
    Topic & Task  & Reference & \multicolumn{1}{l}{Method} \bigstrut\\
    \hline
    \multirow{21}[4]{*}{Clinical Simulation} & \multirow{11}[2]{*}{Digital Twins} & \citet{walsh2020generating} & \multicolumn{1}{l}{Conditional Restricted Boltzmann Machines } \bigstrut[t]\\
          &       & \citet{bertolini2020modeling} & \multicolumn{1}{l}{Conditional Restricted Boltzmann Machines } \\
          &       & \citet{bertolini2022machine} & \multicolumn{1}{l}{Conditional Restricted Boltzmann Machines } \\
          &       & \citet{choi2017generating} & \multicolumn{1}{l}{Generative Adversarial Networks} \\
          &       & \citet{guan2018generation} & \multicolumn{1}{l}{Generative Adversarial Networks} \\
          &       & \citet{cui2020conan} & \multicolumn{1}{l}{Generative Adversarial Networks} \\
          &       & \citet{baowaly2019synthesizing} & \multicolumn{1}{l}{Wasserstein Generative Adversarial Networks} \\
          &       & \citet{zhang2020ensuring} & \multicolumn{1}{l}{Wasserstein Generative Adversarial Networks} \\
          &       & \citet{biswal2020eva} & \multicolumn{1}{l}{Variational Auto-encoders} \\
          &       & \citet{lee2020generating} & \multicolumn{1}{l}{Auto-encoders and Generative Adversarial Networks} \\
          &       & \citet{xu2019modeling} & \multicolumn{1}{l}{Conditional Generative Adversarial Networks} \bigstrut[b]\\
\cline{2-4}          & \multirow{10}[2]{*}{Biosimulation} & \citet{eddy2003archimedes} & \multicolumn{1}{l}{Differential Equations} \bigstrut[t]\\
          &       & \citet{kovatchev2009silico} & \multicolumn{1}{l}{Differential Equations} \\
          &       & \citet{man2014uva} & \multicolumn{1}{l}{Differential Equations} \\
          &       & \citet{gillette2021framework} & \multicolumn{1}{l}{Differential Equations} \\
          &       & \citet{compte2009blood} & \multicolumn{1}{l}{Differential Equations} \\
          &       & \citet{baillargeon2014living} & \multicolumn{1}{l}{Differential Equations} \\
          &       & \citet{zhang2018advances} & \multicolumn{1}{l}{3D Printing} \\
          &       & \citet{low2021organs} & \multicolumn{1}{l}{3D Printing} \\
          &       & \citet{herland2020quantitative} & \multicolumn{1}{l}{3D Printing} \\
          &       & \citet{yu2021reinforcement} & \multicolumn{1}{l}{Reinforcement Learning} \bigstrut[b]\\
    \hline
    \multirow{19}[4]{*}{Individualized Predictive Modeling} & \multirow{14}[2]{*}{Risk Detection} & \citet{rajpurkar2020evaluation} & \multicolumn{1}{l}{Gradient-boosted Decision Trees} \bigstrut[t]\\
          &       & \citet{wang2021survtrace}  & \multicolumn{1}{l}{Transformers} \\
          &       & \citet{wang2022transtab} & \multicolumn{1}{l}{Transformers} \\
          &       & \citet{baumel2018multi} & \multicolumn{1}{l}{Convolutional Networks} \\
          &       & \citet{cheng2016risk} & \multicolumn{1}{l}{Temporal Convolutional Networks} \\
          &       & \citet{ma2018health} & \multicolumn{1}{l}{Temporal Convolutional Networks} \\
          &       & \citet{kam2017learning} & \multicolumn{1}{l}{Multi-layer Perceptron} \\
          &       & \citet{choi2018mime} & \multicolumn{1}{l}{Multi-layer Perceptron} \\
          &       & \citet{rajkomar2018scalable} & Recurrent Neural Networks \\
          &       & \citet{che2017rnn} & Recurrent Neural Networks \\
          &       & \citet{xu2018raim} & Recurrent Neural Networks \\
          &       & \citet{gao2020stagenet} & Recurrent Neural Networks and Convolutional Networks \\
          &       & \citet{wang2021online} & \multicolumn{1}{l}{Graph Neural Networks} \bigstrut[b]\\
\cline{2-4}          & \multirow{5}[2]{*}{Sequential Event Prediction} & \citet{choi2016doctor} & Recurrent Neural Networks \bigstrut[t]\\
          &       & \citet{pham2017predicting} & Recurrent Neural Networks \\
          &       & \citet{ma2017dipole} & Recurrent Neural Networks \\
          &       & \citet{ma2020concare} & Recurrent Neural Networks \\
          &       & \citet{liu2022catnet} & Transformers \bigstrut[b]\\
    \hline
    \multirow{19}[8]{*}{Computer-aided Trial Design} & \multirow{6}[2]{*}{Trial Semantic Parsing} & \citet{lehman2019inferring} & Recurrent Neural Networks \bigstrut[t]\\
          &       & \citet{marshall2017automating} & Recurrent Neural Networks and Language Models \\
          &       & \citet{nye2020trialstreamer} & Recurrent Neural Networks and Language Models \\
          &       & \citet{wang2022trial2vec} & Language Models \\
          &       & \citet{wang2020text} & Sequence-to-sequence Networks \\
          &       & \citet{yu2020dataset} & Sequence-to-sequence Networks \bigstrut[b]\\
\cline{2-4}          & \multirow{4}[2]{*}{Trial Outcome Prediction} & \citet{gayvert2016data} & Random Forests \bigstrut[t]\\
          &       & \citet{hong2020predicting} & \multicolumn{1}{l}{Gradient-boosted Decision Trees} \\
          &       & \citet{qi2019predicting} & Recurrent Neural Networks \\
          &       & \citet{siah2021predicting} & Gradient-boosted Decision Trees \\
          &       & \citet{fu2022hint} & Graph Neural Networks and Language Models \bigstrut[b]\\
\cline{2-4}          & \multirow{4}[2]{*}{Trial Protocol Evaluation} & \citet{liu2021scalable} & Language Models \bigstrut[t]\\
          &       & \citet{liu2021evaluating} & Shapley Scores and Cox Proportional Hazard Model \\
          &       & \citet{xu2021machine} & Poisson Factor Analysis \\
          &       & \citet{rogers2022leveraging} & K-means Clustering and Cox Proportional Hazard Model \bigstrut[b]\\
    \hline
    \end{tabular}
    }
  \label{tab:paper_list}%
\end{table*}%

\section{Clinical Simulation}
\label{sec:simulation}
One of the most expensive aspects of clinical trials is testing. It usually includes in vivo experiments like whole microscopic human tissue cultures, which may raise ethical issues. In silico trials are an ongoing research area aimed at using simulations or machine models to replicate some, if not all, parts of a regular clinical trial.

\subsection{Digital Twins}
\label{sec:digital_twins}
\begin{figure}[h]
\centering
\includegraphics[trim={6cm 3cm 6cm 3cm},clip, width=0.48\textwidth]{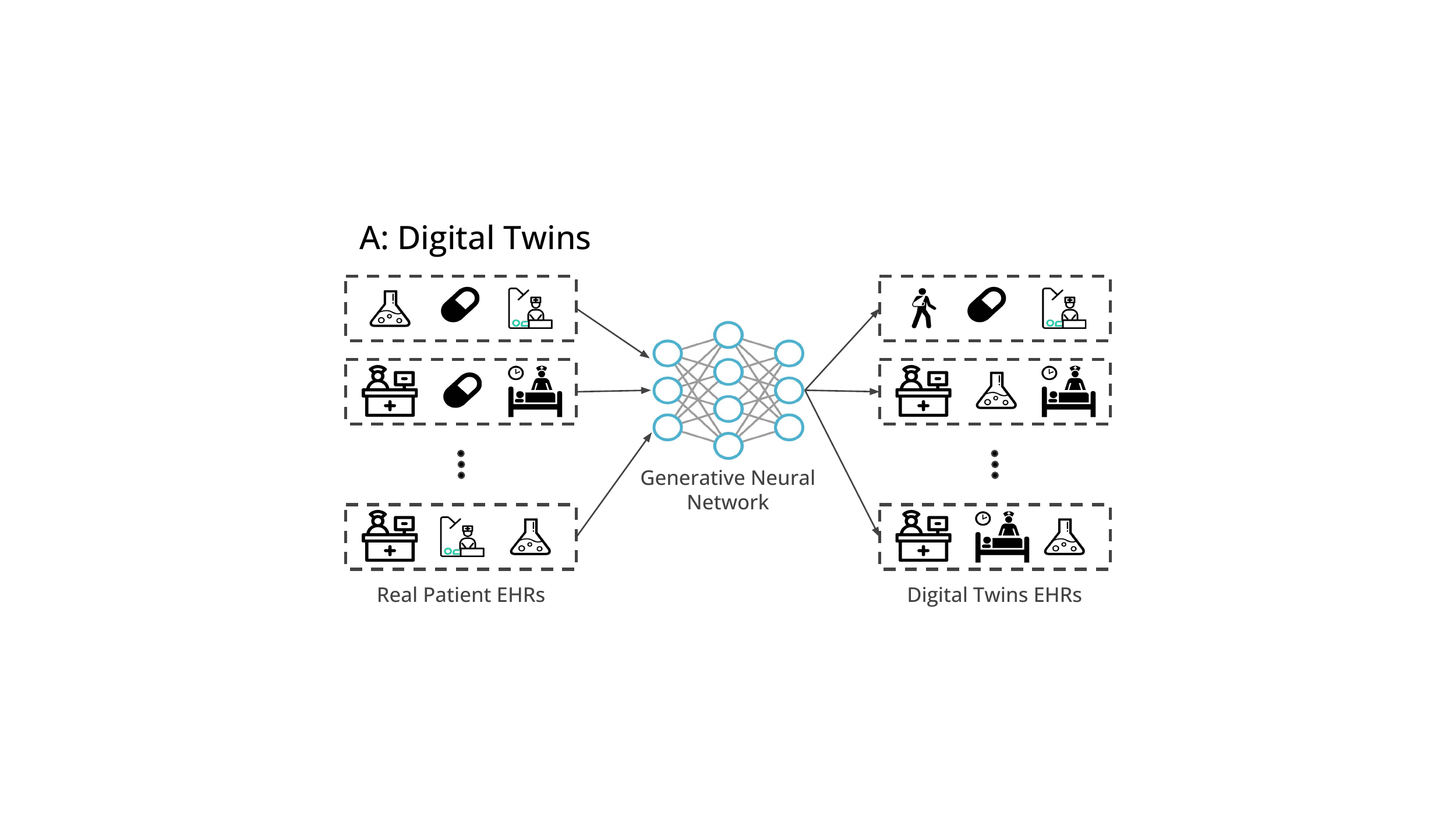}
\includegraphics[trim={6.5cm 4.5cm 6.5cm 4.5cm},clip, width=0.4\textwidth]{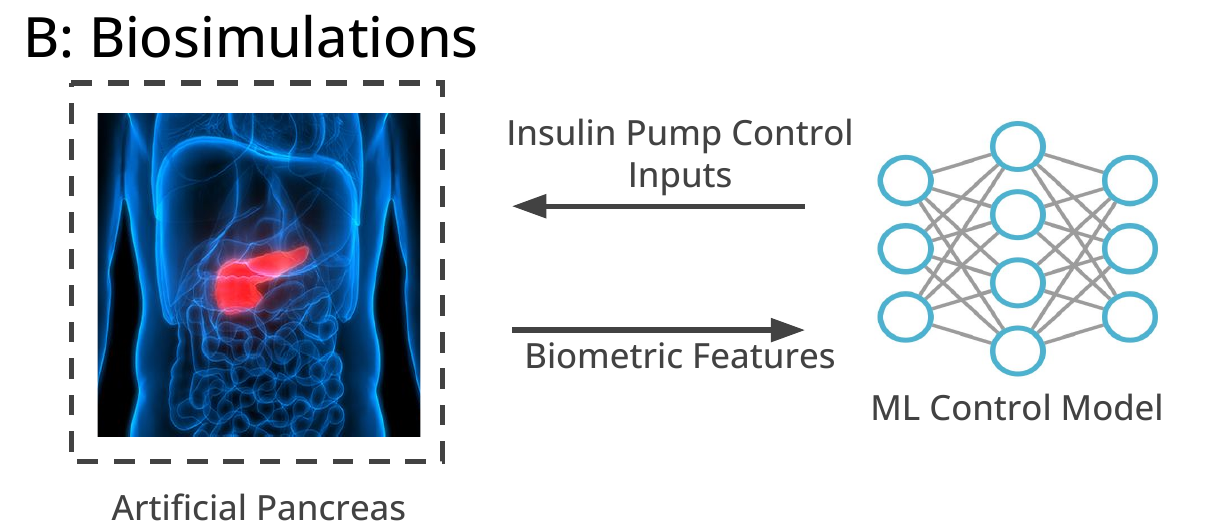}
\caption{
Part A shows an example of Digital Twins: real patient data is used to train a generative neural network for simulating in silico patients (digital twins). See \S \ref{sec:digital_twins}. Part B shows an example ML model learning a control algorithm for diabetes by interacting with a biosimulator such as an artificial pancreas, see \S \ref{sec:Biosimulations} for details.  \label{fig:fig_sec2_digital_twins}}
\end{figure}

One of the primary benefits of in silico clinical trials is the capacity to run experiments in a virtual environment without the need for expensive and time-consuming real-world experiments. One technology being heavily discussed in the clinical trial space is the idea of a digital twin -- an artificial physiological model of a patient that responds to various clinical trial experiments \cite{tao2019digital, laubenbacher2022building}.

The first concept of a digital twin was introduced by National Aeronautics and Space Administration (NASA) through their Apollo 13 mission, which used a simulation of the actual spacecraft to test navigational strategies, inspiring the concept of a "twin strategy" \cite{ferguson_2020}. This allowed NASA to guide the astronauts to safety despite the crisis. However, digital twins are a relatively novel idea in the clinical space. While there has been much work in the form of position papers promoting the general concept of digital twins \cite{fagherazzi2020deep, barricelli2019survey, voigt2021digital, corral2020digital, laubenbacher2022building}, only a few have described their methodology in detail.

A concrete example includes the use of digital twins to model the progression of Multiple Sclerosis (MS) by \citet{walsh2020generating}, which utilizes Conditional Restricted Boltzmann Machines (CRBMs) \cite{hinton2012practical}. 
They showed that digital subjects generated by the CRBM were statistically indistinguishable from actual subjects enrolled in the placebo arms of clinical trials for MS. Similarly, \citet{bertolini2020modeling} described a generative model composed of two CRBMs that accurately simulated subjects with Mild Cognitive Impairment (MCI) and Alzheimer’s Disease (AD). They showed that linear classifiers were unable to distinguish actual subjects from their digital twins based on the generated features; \citet{bertolini2022machine} described a similar technique on the Amyotrophic Lateral Sclerosis (ALS) that demonstrated how integrating CRBM digital twins into clinical trial designs could reduce the sample size required to achieve the desired power. Fig. \ref{fig:fig_sec2_digital_twins} illustrates how digital twins are created from subjects.

Another closely related research area is synthetic patient data generation. This research area seeks to simulate patient records, usually for specific disease conditions. For example, one can predict future patient EHR data using current patient baseline metrics or historical EHR data. 
Early works on patient record generation methods used rule-based methods \cite{lombardo2008ta, buczak2010data, mclachlan2016using}. 
However, the rule-based methods cannot provide realistic and complex patient data to support general machine learning (ML) tasks.
In contrast, deep generative models advanced by the power of neural networks, e.g., variational auto-encoders (VAE) \cite{kingma2013auto}, generative adversarial networks (GAN) \cite{goodfellow2014generative}, gain attention of researchers. \citet{choi2017generating} introduced MedGAN, pioneering GANs for discrete patient records generation. Multiple works subsequently worked to improve GANs for EHRs generation. \citet{guan2018generation} introduced GANs for generating medical texts instead of just structured EHR data. \citet{baowaly2019synthesizing, zhang2020ensuring} tested multiple GAN architectures and loss functions, like Wasserstein GAN with gradient penalty \cite{gulrajani2017improved} and boundary-seeking GAN \cite{hjelm2017boundary}. Other works used other generative models such as VAE, which can produce realistic data that improve downstream model accuracy when added to the training set \cite{biswal2020eva}. In some scenarios, a hybrid of GANs and other models can yield superior performances. For example, \citet{lee2020generating} proposed dual adversarial autoencoder (DAAE), which learns set-valued sequences of medical entities, by combining a recurrent autoencoder with two GANs, adversarially learning both the continuous latent distribution and the discrete data distribution. This is to address the issue of mode collapse in GANs, where GANs only generate a portion of the valid possible data. \citet{cui2020conan} proposed Complementary pattern Augmentation (CONAN), which uses GANs, adversarial training, and max-margin classification to improve rare disease detection from EHR data. \citet{Biswal2020-cp} proposes a multi-modal GAN-based generation method with self-supervised learning to generate both X-ray images and associated clinical notes. 

\textit{Machine learning formulation.} For unconditional generation (not specifying any disease condition), the ML model tries to generate patient records from scratch, usually taking random noise as inputs. For conditional generation (specific disease conditions are given), the inputs can be the historical records or key disease phenotypes in the form of medical codes. The generator will try to produce patient records that match the specified input condition. More formally, let  $h^k$ be the EHR for patient $k$. An EHR generation task or a digital twin cohort may be specified as generating a synthetic patient based on generative model $p()$ trained on real patients $h^1$ to $h^k$: $h^{synthetic} \sim p(h^{synthetic} | h^1, h^2, \dots h^k)$.

\textit{Data sources.} Real EHRs are available from \href{https://physionet.org/content/mimiciii/1.4/}{MIMIC-III}, \href{https://physionet.org/content/mimiciv/0.4/}{IV}, and \href{https://eicu-crd.mit.edu/gettingstarted/access/}{eICU} for learning AI models that are capable of making synthetic EHRs generation on variant conditions. \href{https://data.projectdatasphere.org/}{Project data sphere} provides access to patient-level trial records for over 200 clinical trials.

\subsection{Biosimulations}
\label{sec:Biosimulations}
Mathematical or physical simulations of physiology are a subarea of simulation. Rather than using ML methods, these approaches focus on systems of differential equations, physics  simulations, and other traditional methods such as systems of equations. Biosimulations can have several benefits, such as being more interpretable than black box ML methods. However, ML still has a large part to play in this area. Since many of the aforementioned approaches rely on concrete, mathematical and physical simulations of physiology, ML can help tune model parameters for better accuracy. For example, in the case of virtual trials for glycaemic control, there are still parameters that need to be inferred for both the control algorithm and the simulation model. Black box ML models such as Bayesian optimization may be helpful in such cases.

For example, the artificial pancreas is a popular area of research for in silico closed-loop control of blood glucose in diabetes. Usually, the process consists of a glucose sensor and a control algorithm for the delivery of insulin \cite{cobelli2011artificial}. Another example of this is Archimedes, which is a trial-validated model of diabetes using differential equations of more than 100 interacting features \cite{eddy2003archimedes}. Independent validation on 18 other real-world trials showed that Archimedes is a realistic representation of the anatomy and physiological responses of diabetes \cite{eddy2003validation}. \citet{compte2009blood} designed an insulin controller developed it on in silico trial simulations on a mathematical system model \cite{chase2007model}. These models have been validated using real-world retrospective hospital control data \cite{chase2010validation}. Other relevant works on artificial pancreas include an FDA-approved computer simulation of glucose sensing, and insulin delivery \cite{kovatchev2009silico}, improvements on existing simulators \cite{man2014uva}, and improvements on the control algorithm using modern ML methods like reinforcement learning \cite{yu2021reinforcement}. Furthermore, virtual trials for glycaemic control have also been shown to be generalizable across hospitals worldwide \cite{dickson2017generalisability}.

Other system models include mathematical models of the heart  fitted to simulate cardiac electrophysiology from ECGs time-series \cite{gillette2021framework}; modeling the dynamics of human liver \cite{cook2015systems}, a simulator for cardiac excitation and contraction in the human heart via a classical resistance-based Windkessel model \cite{westerhof2009arterial} called the the Living Heart Project \cite{baillargeon2014living}. 


\textit{Machine learning formulation.} To use ML models to develop control algorithms for the biosimulators, techniques like reinforcement learning may be used. For example, the input observations at state $s \in S$ could be the state of the organ, given by its features such as glucose level; the ML model output would be an action $a \in A$ to control biosimulator. For example, the available actions could include how much insulin to give. Let $r$ denote the reward corresponding to the patient's health status.

More formally, a given trajectory would look like: $$s_0, a_0, r_0, s_1, a_1, r_1, s_2, a_2, r_2, \dots$$
Let $\gamma$ be the discount. We would then like to maximize the expected discounted return at time $t$: $${G_t} \Dot{=} \sum_{k=0}^\infty \gamma^k r_{t+k+1}$$

Solving this optimization will yield a learned optimal policy function $\pi(s) \xrightarrow{} a$ that learns what actions to take at a given state.

\textit{Data sources.} 
In this case, the biosimulations themselves often act as a simulation environment to enable downstream machine learning tasks. Archimedes \cite{eddy2003archimedes} is one source of a trial-validated model of diabetes, which may produce pathophysiology of diabetes at a high level of clinical detail. The Living Heart Project \cite{baillargeon2014living} is another source of data from an integrative, predictive model of the living human heart. Both simulators are validated by selected clinical trials \cite{chase2010validation} or by clinical observations from other studies \cite{wong2013computational}. 

\section{Individualized Predictive Modeling}
\label{sec:prediction}
\begin{figure*}[t]
\centering
\includegraphics[width=1.0\textwidth]{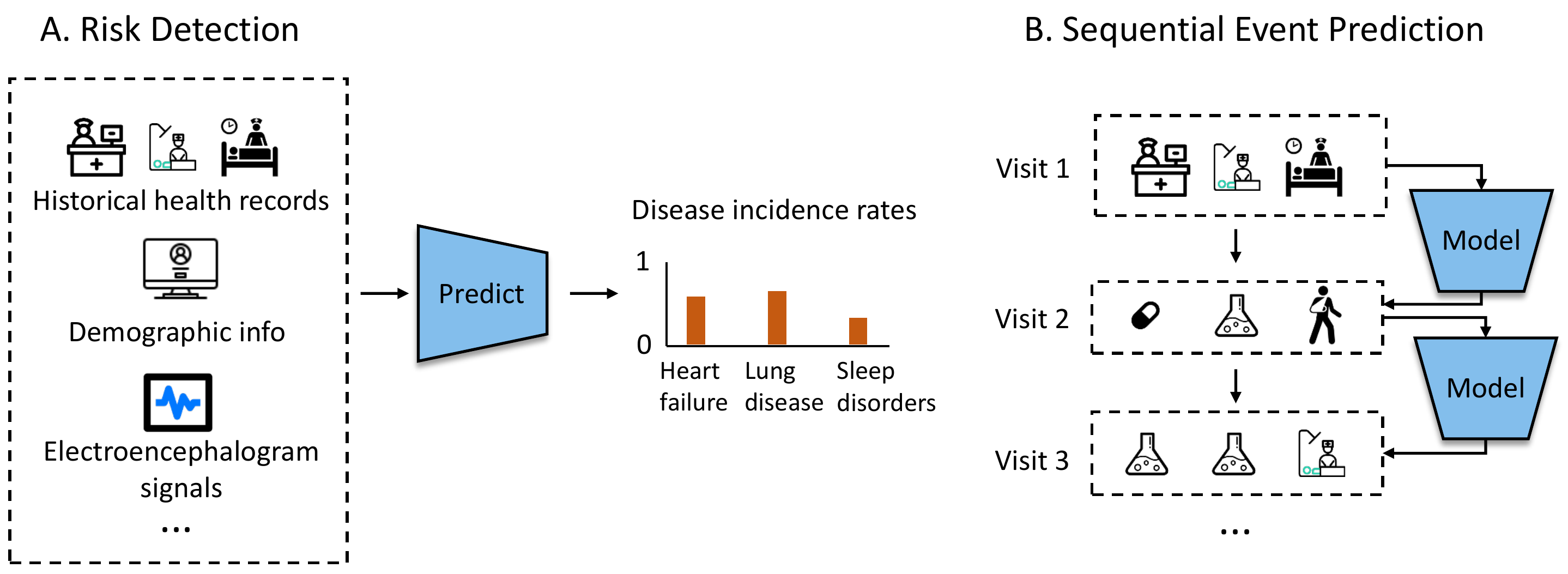}
\caption{Task illustration for the topic ``Individualized Predictive Modeling''. (A) Multi-modal data collected from various sources is aggregated as the inputs for predictive models for predefined targets, e.g., mortality or disease onset (\S \ref{sec:risk_detect}); (B) A model iteratively predicts the events happened in the next visit taking historical visits as inputs. (\S \ref{sec:sequential}). \label{fig:fig_sec3}}
\end{figure*}

During clinical trials, predictive modeling catches abnormalities, e.g., alarming responses or patient in-adherence, that would suggest the alteration of predefined treatment options. Additionally, depending on the data, early termination of the trial should be considered if the interventions appear harmful. If conclusive results are obtained, the trial may be stopped early and moved onto the next appropriate stage \cite[Chapter ~16]{friedman2015fundamentals}. Moreover, traditional trials only verify the average efficacy and safety among trial participants rather than individual subjects. They do not adapt to the characteristics of participants. The emergence of AI can potentially address this personalization challenge through patient predictive modeling. This section reviews how AI can mine through patient records and identify patterns or make individualized predictions for response metrics, as shown in Fig. \ref{fig:fig_sec3}.

In clinical trials, AI has achieved individualized patient outcome predictions for several diseases. For example, \citet{rajpurkar2020evaluation} leveraged gradient boosted decision trees to predict the improvement of depressive symptoms for patients receiving antidepressant treatment based on pre-treatment symptom sc\-ores and electroencephalographic measures; \citet{hong2020predicting} predicted clinical drug toxicity with drug properties and target property features through an ensemble of classifiers; \citet{de2021towards} built a model to predict drug response for neurological disorder patients integrated with genetics data; \citet{wang2021survtrace} modeled and predicted the survival rate of breast oncology patients using transformers \cite{vaswani2017attention}; \citet{wang2022transtab} proposed a transferable transformer model that learned from data across multiple oncology trials to boost mortality predictions for each trial.

The literature on outcome prediction using EHR data has also been fruitful. EHRs usually contain demographic information, e.g., age and gender, and a sequence of clinical visits, e.g., diagnoses, procedures, lab tests, and medication prescriptions. Two main categories of predictive modeling tasks are pertinent to the case: risk detection and sequential prediction of clinical events \cite{xiao2018opportunities}.

\subsection{Risk Detection} \label{sec:risk_detect}
Risk detection seeks to predict the risk of the target event, e.g., disease onset, taking EHRs as inputs. The risk detection is formed in three types: First, as either binary classification (e.g., onset of a specific disease \cite{cheng2016risk, kam2017learning,Choi2017-os} or mortality \cite{rajkomar2018scalable}); Second, as multi-class classification (e.g., disease stage prediction \cite{che2017rnn} and  disease categories \cite{wang2021online}); Third, as multi-label classification (e.g., diagnosis code assignment \cite{baumel2018multi}). Longitudinal patient data such as EHRs are often used as input for risk detection. For example, \citet{kam2017learning} applied long short term memory (LSTM) model to the vital signals drawn from the MIMIC-III data \cite{johnson2016mimic} for sepsis detection; \citet{ma2018health} designed a hybrid Recurrent Neural Network (RNN) and Convolutional Neural Network (CNN) model that modeled heterogeneous EHRs to make health risk predictions; \citet{choi2018mime} learned embeddings based on the multilevel hierarchy of medical codes in EHRs to make heart failure predictions; \citet{xu2018raim} processed multimodal physiological monitoring data, e.g., heart rate, pulse, electrocardiogram, then fused them with clinical events as input for LSTM to predict decompensation risk and length of ICU stay; \citet{gao2020stagenet} accounted for disease progression stage using an LSTM and CNN to make decompensation risk and mortality risk prediction for the end-stage renal disease. 

\textit{Machine learning formulation.} The formulation is as follows: specify the target response to be predicted, e.g., mortality, readmission, length of stay, transform the input EHRs to either static descriptive features or sequential event features, then predict the target using the EHR features.

\textit{Data sources.} \href{https://physionet.org/content/mimiciii/1.4/}{MIMIC-III} and \href{https://physionet.org/content/mimiciv/0.4/}{IV} are the two public datasets that offer comprehensive EHRs from ICU stays. Several risk prediction tasks can be built, including decompensation prediction, mortality prediction, and readmission prediction.  \href{https://data.projectdatasphere.org/}{Project data sphere} provides access to patient-level trial records for over 200 clinical trials upon application. \href{https://seer.cancer.gov/data-software/}{SEER} offers oncology incidence data across the United States, which can be used for mortality risk prediction and time-to-event prediction.

\subsection{Sequential Event Prediction}\label{sec:sequential}
Besides detecting specific diseases, we also observe AI's use for predicting other events from longitudinal EHR data. While some of the disease detection works are capable of predicting events like readmission, they need to be performed in two separate tasks, i.e., one model for disease detection and the other for readmission prediction \cite{rajkomar2018scalable}. In contrast, sequential event prediction seeks to make a multi-label classification for multiple events at once. For instance, \citet{choi2016doctor} adopt recurrent neural networks (RNNs) on longitudinal EHRs to make multi-label predictions on diagnosis, procedures, and medications; \citet{pham2017predicting} leveraged LSTM to model the longitudinal illness state to predict disease progression in time order; \citet{ma2017dipole} adapted attention mechanisms and RNNs for interpretable medical codes prediction; \citet{liu2022catnet} treated input EHRs as multimodal time series, building a time-aware cross-event attention model to capture the correlations among events and unify the modeling of heterogeneous EHRs (including medication, diagnosis, procedure, and lab tests).

\textit{Machine learning formulation.} The inputs are the historical events (longitudinal EHR); the output consists of stepwise, multi-label predictions for multiple target events. 

\textit{Data sources.} 
The most common data for sequential prediction are longitudinal EHR data with multiple visits per patient. For example,
\href{https://physionet.org/content/mimiciii/1.4/}{MIMIC-III} and \href{https://physionet.org/content/mimiciv/0.4/}{IV} provide real longitudinal EHRs that fit the sequential event prediction task, e.g., automatic coding and drug recommendation. Synthetic EHRs like \href{https://www.cms.gov/Research-Statistics-Data-and-Systems/Downloadable-Public-Use-Files/SynPUFs}{SynPUFs} is another dataset commonly used for testing AI models for sequential prediction.  \href{https://data.projectdatasphere.org/}{Project data sphere} contains sequential clinical trial data that can be used to predict trial adverse events and mortality.

\begin{figure*}[t]
\centering
\includegraphics[width=1.0\textwidth]{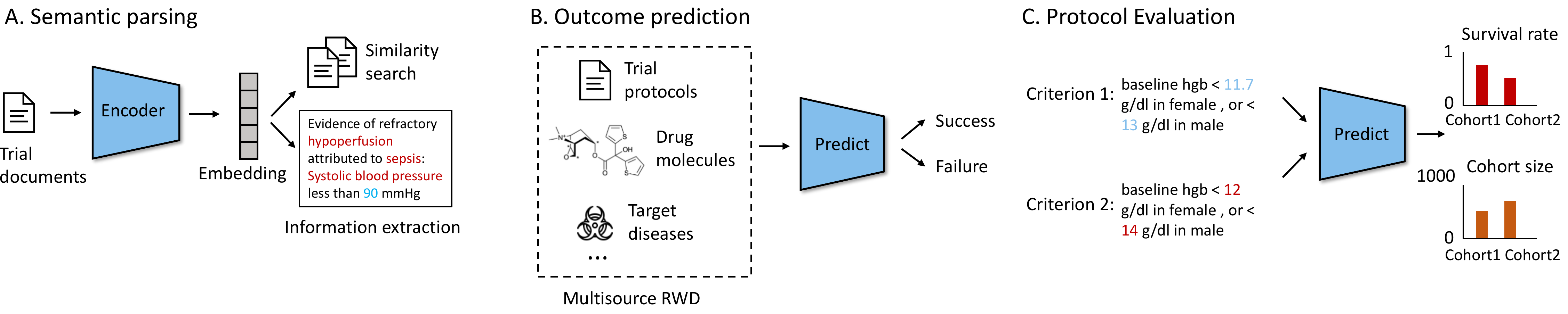}
\caption{Task illustration for the topic ``computer-aided trial design''. (A) An encoder encodes trial documents to embeddings, which are used for trial retrieval or extraction of key entities (\S \ref{sec:semantic_parsing}); (B) A model predicts whether a clinical trial will succeed or fail given multisource real-world data (\S \ref{sec:outcome}); (C) Raw trial protocols, e.g., criteria, are parsed. A predictor is used to evaluate how primary metrics change if any trial elements change (\S \ref{sec:evaluation}).  
\label{fig:fig_sec4}}
\end{figure*}

\section{Computer-aided Trial Design}
\label{sec:design}
The clinical trial design plays a central role in the success of the clinical trial. Researchers need to design trial protocols to answer scientific questions, including the eligibility criteria, population size, duration, etc. Historically, the trial design completely relies on the experiences of clinical trial planners. As a result, some protocols like eligibility criteria are far from optimal and may harm trial success \cite{kim2015modernizing}. By contrast, AI algorithms may help automated trial design, as AI has been demonstrated in drug discovery such as molecule design \cite{jin2018junction}. As such, AI for trial protocol optimization has been quickly growing in recent years \cite{woo2019ai}. AI  models can extract structured information from trial documents \cite{yu2020dataset}, predict the trial outcomes with multimodal trial-related real-world data (RWD) \cite{fu2022hint}, and evaluate how each component of the trial protocol influences the trial outcomes \cite{liu2021evaluating}.  Fig. \ref{fig:fig_sec4} shows these examples.

\subsection{Trial semantic parsing}\label{sec:semantic_parsing}
Real-world data such as clinical trial protocols or scientific publications about clinical trial results are crucial for facilitating trial design. However, trial records are semi-structured documents with both unstructured texts and structured result tables. Most AI models cannot handle such complex semi-structured data before they are transformed into structured forms such as SQL queries. For instance, \citet{wang2020text} extracted medical queries from MIMIC-III and generated SQL using neural networks; \citet{yu2020dataset} focused on SQL generation for clinical eligibility criteria. They introduce a new dataset that includes the first-of-its-kind eligibility-criteria corpus and the corresponding queries. It enables training models that translate the eligibility criteria to executable SQL queries.

Evidence-based medicine underpins new trial design decisions, but clinical trial results are disseminated as scientific papers in free texts. One way to extract relevant information from these free texts is through evidence synthesis, which automatically extracts Population, Interventions/Comparators, and Outcomes (PICO) sentences from the scientific articles. For example, \citet{marshall2017automating} developed RobotViewer, a model that integrates ML and NLP to extract texts describing key trial characteristics  as summary reports; \citet{lehman2019inferring} used a suite of ML models to extract the reported outcomes and infer the  treatment efficacy from scientific corpus; \citet{nye2020trialstreamer} built RobotViewer to support the synthesis of evidence maps extracted from multiple trials. 

Another line of research aims to parse and formalize free-text eligibility criteria so that the logic and properties represented by trial criteria can be utilized to match patients to trials \cite{kang2017eliie}. By parsing and extracting evidence from trial documents, one can structure trials by standard concepts and the corresponding values, which paves the way for predictive AI, e.g., tabular learning \cite{wang2022transtab}, and document embedding \cite{wang2022trial2vec}. A related task is to develop ML models to match patient records to trial eligibility criteria descriptions automatically~\cite{Gligorijevic2019-ke, Zhang2020-ku, Gao2020-zh}.

\textit{Machine learning formulation.} Two machine learning tasks are associated with this scenario: (1) Given a free-text or semi-structured trial document, extract the key entities (e.g., PICO) which capture the essential information in a clinical trial; (2) Given trial eligibility criteria represented by free texts, formalize each criterion to be machine-readable logics, e.g., SQL; (3) Given patient records and free text eligibility criteria, automatically find the patients that match the eligibility criteria.

\textit{Data sources.} Raw trial eligibility criteria are available on \href{https://clinicaltrials.gov/}{ClinicalTrials.gov}. \citet{yu2020dataset} released Criteria2SQL, an annotated dataset with paired eligibility criteria and SQL queries, which makes it possible to train AI models to translate from natural language criteria to the corresponding logic forms.

\subsection{Trial outcome prediction}\label{sec:outcome}
The cost for bringing a molecule to market can take billions of dollars \cite{martin2017much}. The process can take many years and yield a success rate of as low as 13\% from phase I to market \cite{blass2015basic}. Predicting the trial outcomes is thus crucial for the economic consideration of drug development. We can avoid running clinical trials that are highly probable to fail. Stakeholders can save unnecessary funding/time from those trials that doom to fail, reinvesting in other promising trials or re-designing trials to improve the chance of success. 

The emergence of AI sheds light on leveraging multi-modal RWD for accurate trial outcome predictions at the planning stage. Practitioners can iterate the trial plan referring to the predictions made by AI until no better change can be made. \citet{gayvert2016data} used chemical structure and properties of drugs and targets and a random forest model to predict the toxicity of clinical drugs; \citet{qi2019predicting} employed RNNs to model both static and time-variant variables to predict the phase III trial outcomes using patient records from previous phases; \citet{siah2021predicting} predicted the drug approval on both the drug features and trial features using statistic ML methods; \citet{fu2022hint} further fused multi-source RWD, e.g., drug molecules, disease ontology and trial eligibility criteria, to support outcome predictions for all phases of trials. 

\textit{Machine learning formulation.} Given trial design protocols and the corresponding multi-modal RWD, e.g., molecule structures, drug-drug interaction graphs, and disease ontologies, predict the outcomes for trials from all phases. 

\textit{Data sources.} \href{https://clinicaltrials.gov/}{ClinicalTrials.gov} is a comprehensive open database that contains clinical trial protocols and their status. There are over 400K research studies in total. Researchers may process the raw protocols and formulate a supervised trial outcome prediction task of the trial status as the target. Additional RWD including drug molecules (e.g., \href{https://go.drugbank.com/}{DrugBank} and \href{https://moleculenet.org/}{MoleculeNet}) and ontologies like ICD codes (\href{https://www.cms.gov/Medicare/Coding/ICD10}{ICD-10}), drugs (\href{https://www.who.int/tools/atc-ddd-toolkit/atc-classification}{ATC codes}), and diseases (\href{https://www.hcup-us.ahrq.gov/toolssoftware/ccs10/ccs10.jsp}{CCS-10}) are beneficial to make better trial outcome predictions. \citet{siah2021predicting} created a challenge for drug approval prediction which includes 10K phase II clinical trials. \citet{fu2022hint} created a trial outcome prediction benchmark which includes many small molecule trials.

\subsection{Trial protocol evaluation}\label{sec:evaluation}
Trial outcome predictors provide an overall success probability, given clinical trial information. However, how exactly to improve these trials is still an elusive question. In general, researchers may dive into eligibility criteria to investigate if they are too stringent to recruit sufficient patients  \cite{van2007eligibility} or too relaxed, reducing toxicity and adverse events \cite{kim2017broadening}. This area of research yields a plethora of works in ML for data-driven eligibility criteria design. For instance, \citet{liu2021scalable} extracted key entities from trial criteria using NLP to match patients records to expand the potential cohort size; \citet{xu2021machine} derived subgroups of patient cohort with clustering to identify the ones with less severe adverse events, which can inform the design of criteria; \citet{kim2021towards} manually adjusted criteria to broaden patient recruitment and assess the criteria influence on the inclusion and exclusion groups; \citet{liu2021evaluating} utilized Shapley scores \cite{lundberg2017unified} to estimate the change of hazard ratio of the included oncology patients when removing each individual criterion; \citet{rogers2022leveraging} simulated a set of criteria and assessed the correspondingly retrieved cohort in terms of the hazard ratio, finding that it is possible to identify a group of superior criteria combinations via k-means clustering on the count of cohort and hazard ratio.

Existing methods quantitatively estimate the contribution of individual criteria to the trial generalizability or further consider the adverse events and hazard rate of the included cohort. Modifications such as removing one criterion can be made based on the analysis. However, they cannot capture the correlations between criteria, i.e., the effect of removing criterion A and B simultaneously usually does not equal the summation of independently removing A and B, respectively. 

\textit{Machine learning formulation.} 
Given a set of trial eligibility criteria $\{e_1,e_2,\dots,e_m\}$ and a group of candidate participants $\{x_1,\dots,x_n\}$ where $x$ represents the individual historical healthcare records in table or in sequence form, estimate the effect to the target of interest by varying the matched cohort size $S$, removing criterion $e_1$ or modifying $e_1$ to $\tilde{e}_1$, etc.

\textit{Data sources.} Researchers may match trial records on \href{https://data.projectdatasphere.org/}{Project data sphere} to the trials on \href{https://clinicaltrials.gov/}{ClinicalTrials.gov} to retrieve the corresponding trial designs. Simulation is then feasibly used to evaluate the trial protocols.

\section{Discussion \& Conclusion}
In silico trials are virtual trials that conduct all and parts of clinical trials digitally. Thanks to the potential economic benefit, researchers and practitioners are excited about the applications of in silico trials. This paper reviews different topics related to in silico trials, including clinical simulation \S\ref{sec:simulation}, individualized predictive modeling \S\ref{sec:prediction}, and computer-aided trial design \S\ref{sec:design}.
Next, we summarize the challenges and opportunities of AI for in silico trials:

\begin{itemize}
    \item \textbf{Lack of large-scale datasets and benchmarks}. It is well known that the current AI models rely heavily on high-quality labeled training data. There exist limited benchmark data for trial-related tasks. Only a small fraction of clinical trial records are available for application. Creating and releasing benchmark data for various in silico trial tasks can encourage further work in the domain. The only noticeable exception is the TOP benchmark for trial outcome prediction~\cite{fu2022hint}. 
    
    \item \textbf{Out-of-distribution (OOD) prediction}. Most clinical trials are performed for common diseases, and these AI models often cannot generalize to rare diseases due to data distribution shifts and small sample sizes. It is challenging to achieve the descent OOD performance for new unseen trials using AI models.
    \item \textbf{Inference in low-data scenarios}. Despite the current success of AI models in healthcare, AI for clinical trials often suffer from small sample sizes. Most clinical trials have less than a thousand participants, with most phase I trial having less than a hundred participants. This poses a major challenge for learning powerful AI models for clinical trials. Transfer learning is a promising direction to allow AI models to apply across clinical trials.
    \item \textbf{Federated learning across organizations}. 
    One approach to tackling the small sample size problem in a single clinical trial is to train AI models using data from multiple trials. However, most trials have unique designs and collect different measurements inconsistent across trials. Plus, clinical trial data are sensitive and difficult to share across organizations due to privacy and legal constraints. Federated learning provides a promising framework to support AI model building across organizations without direct data sharing with small initial success in healthcare application~\cite{Dayan2021-wh}. 
    However, due to heterogeneity in trial data, federated learning is still impractical to use for clinical trial applications.
\end{itemize}

Finally, in silico trials is an exciting topic for AI researchers. This paper provides a comprehensive review of research on the intersection of AI and in silico trials. We summarized the key topics and presented machine learning formulation and available datasets to consider.
Finally, we discussed the current open challenges of AI and in silico trials that hope to inspire further research. 

\bibliography{ref}
\bibliographystyle{unsrtnat}

\appendix

\end{document}